\documentclass[12pt,dvipdfmx]{article}
\usepackage{graphicx}

\newcommand\pubnumber{NuPhys2016-Nagai}
\newcommand\pubdate{May 1, 2017}

\def\napoli{Department of Physics, University of Colorado Boulder\\
390 UCB, University of Colorado, Boulder, CO 80309, USA}
\def\support{\footnote{Yoshikazu.Nagai@cern.ch}}

\def\Title#1{\begin{center} {\Large #1 } \end{center}}
\def\Author#1{\begin{center}{ \sc #1} \end{center}}
\def\Address#1{\begin{center}{ \it #1} \end{center}}

\newcommand\pubblock{\rightline{\begin{tabular}{l} \pubnumber\\
         \pubdate  \end{tabular}}}
\newenvironment{Abstract}{\begin{quotation}  }{\end{quotation}}
\newenvironment{Presented}{\begin{quotation} \begin{center} 
             PRESENTED AT\end{center}\bigskip 
      \begin{center}\begin{large}}{\end{large}\end{center} \end{quotation}}

\def\beq{\begin{equation}}
\def\eeq#1{\label{#1}\end{equation}}
\def\eeqn{\end{equation}}

\def\beqa{\begin{eqnarray}}
\def\eeqa#1{\label{#1}\end{eqnarray}}
\def\eeqan{\end{eqnarray}}

\let\bar=\overbar

\def\Dslash{\not{\hbox{\kern-4pt $D$}}}
\def\dslash{\not{\hbox{\kern-2pt $\del$}}}

\def\msb{{\bar{\ssstyle M \kern -1pt S}}}

\begin{document}
\begin{titlepage}
\pubblock

\vfill
\Title{Hadron Production Experiments}
\vfill
\Author{ Yoshikazu Nagai\support}
\Address{\napoli}
\vfill
\begin{Abstract} 
Precise prediction of the neutrino flux is a key ingredient to achieving the physics goals of accelerator-based neutrino experiments.
In modern accelerator-based neutrino experiments, neutrino beams are created by colliding protons with a nuclear target.
Secondary hadrons are produced in these collisions, and their decays contribute to the neutrino flux.
The hadron production is the leading systematic uncertainty source on the neutrino flux prediction; therefore its precise measurement is desirable.

In these proceedings, 
review of recent hadron production measurements and the latest results from the NA61/SPS Heavy Ion and Neutrino Experiment (NA61/SHINE) are presented.
In addition, plans of NA61/SHINE hadron production measurements for the next generation neutrino experiments and NA61/SHINE physics program extension beyond 2020 are discussed.
\end{Abstract}
\vfill
\begin{Presented}
NuPhys2016, Prospects in Neutrino Physics\\
Barbican Centre, London, UK,  December 12--14, 2016
\end{Presented}
\vfill
\end{titlepage}
\def\thefootnote{\fnsymbol{footnote}}
\setcounter{footnote}{0}

\section{Introduction} 

In the neutrino oscillation analysis, number of observed neutrino events at the near and far detectors can be written as proportional to the neutrino flux and cross-section:
\begin{eqnarray}
N_{ND}&\propto&\int \Phi_{ND} \cdot \sigma \,\, dE_\nu  \nonumber \\ 
N_{FD}&\propto&\int \Phi_{FD} \cdot \sigma \cdot P_{osc} \,\, dE_\nu  \nonumber  \\
            &\propto&\int R_{\frac{FD}{ND}} \cdot \Phi_{ND} \cdot \sigma \cdot P_{osc} \,\, dE_\nu  \nonumber
\end{eqnarray}
where $\Phi_{ND(FD)}$, $\sigma$, $P_{osc}$, and $R_{\frac{FD}{ND}}$ denote neutrino flux at the near (far) detector,   neutrino cross-section on the target material, the neutrino oscillation probability, and far to near neutrino flux ratio, respectively. 
As indicated in these equations, well-understood neutrino flux is a key ingredient for the neutrino oscillation analysis.

\begin{figure}[htb]
\centering
\includegraphics[height=37mm]{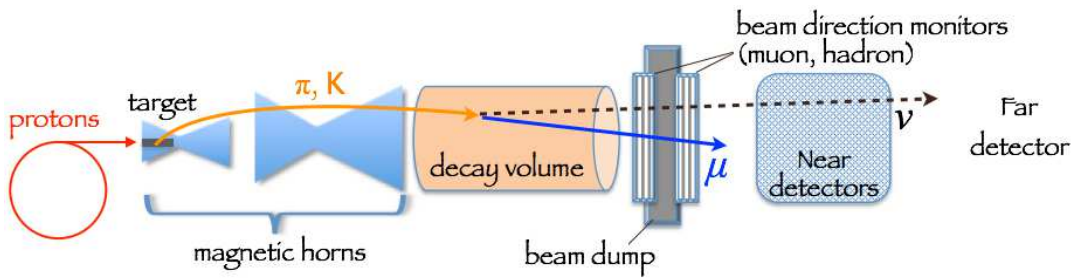}
\caption{Typical beamline design in modern accelerator-based neutrino experiments.}
\label{Fig:Beamline}
\end{figure}

A typical beamline design in modern accelerator-based neutrino experiments is shown in Figure~\ref{Fig:Beamline}.
In modern experiments, neutrino beams are created by colliding protons with a light nuclear target, such as carbon or beryllium targets.
Secondary hadrons are produced via primary interactions of beam protons and their decays contribute to the neutrino flux. 
Some fraction of secondary hadrons re-interacts in target or with other materials out of target. 
These secondary interactions also produce hadrons which contribute to the neutrino flux.
Neutrinos mostly come from decays of charged pions at lower energy region, 
while kaon contributions to the neutrino flux are getting larger at higher energy region.
Therefore, precise hadron production measurements on pions and kaons exiting target are necessary.

In general, beam simulation is used to predict the initial neutrino flux at the near and far detectors.
Since the hadronic interaction in the target and beamline materials is non-perturbative QCD process,
there exists a difficulty of theoretical calculations.
Therefore, hadron production simulation relies on models,
such as FLUKA~\cite{Ferrari:2005zk,FLUKA:2014} or GEANT4~\cite{Agostinelli:2002hh}.
Individual model shows different predictions, resulting in the large systematic uncertainty on the neutrino flux prediction.
Since the hadron production is known as the leading systematic uncertainty source on the neutrino flux prediction,
validations of the hadronic interaction models with precise hadron production measurements are highly desirable.

\section{Hadron production measurements}

In this section,
methodology for hadron production measurements are first discussed.
Then, review of recent hadron production experiments and summary of available datasets are presented.

\begin{figure}[htb]
\centering
\includegraphics[height=45mm]{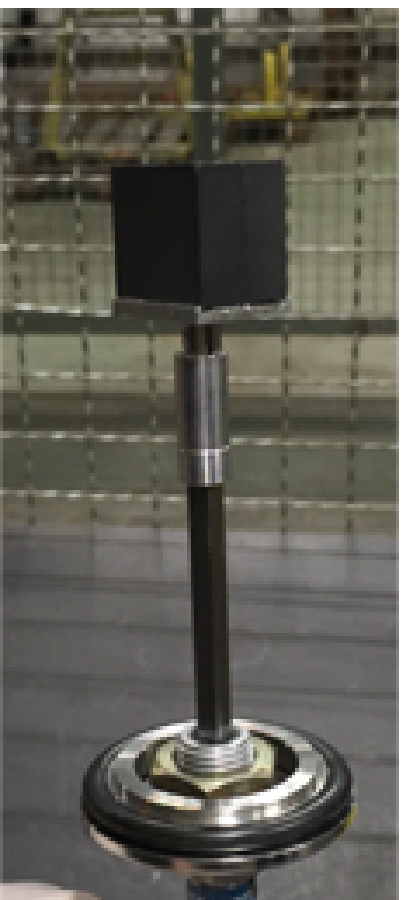} \,\,\,\,\,
\includegraphics[height=45mm]{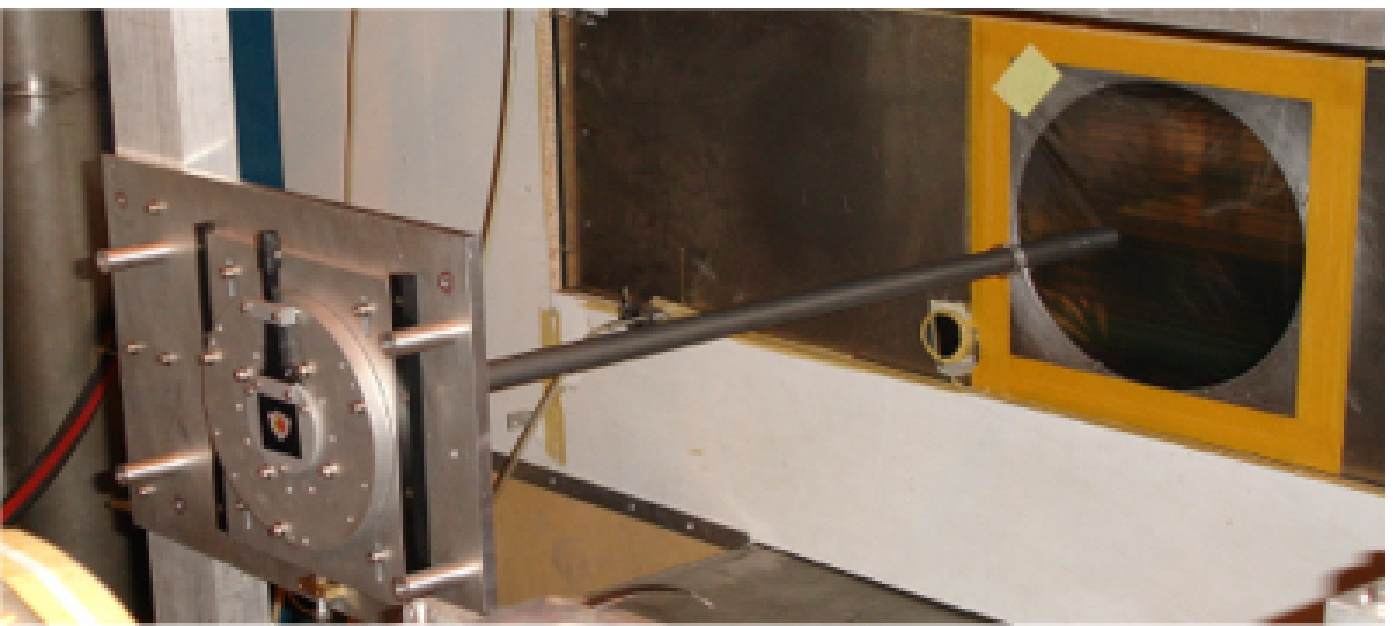}
\caption{(Left) 2\,cm long thin carbon target  in the NA61/SHINE experiment. (Right) 90\,cm long T2K replica target  in the NA61/SHINE experiment.}
\label{Fig:Target}
\end{figure}

Hadron production experiments perform measurements using two types of target: \underline{thin target} and \underline{replica target}.
In Figure~\ref{Fig:Target}, examples of thin and replica target are shown, 
which are used in the NA61/SPS Heavy Ion and Neutrino Experiment (NA61/SHINE).

For the thin target measurement, hadron beams (protons, pions, or kaons) are 
shot on a few \% nuclear interaction length ($\lambda_{\rm int}$) targets.
Main goals are to measure: inelastic cross-section, 
differential cross-section of produced hadrons, and yield of produced hadrons.
The thin target measurements are used to constrain systematic uncertainty on the production of secondary hadrons from the primary interactions. 

Although various thin target measurements are available to constrain primary interactions, some fraction of the neutrino flux comes from decays of hadrons produced through re-interactions of secondary particles.
Only with the thin target measurement, phase space of such production processes are not fully covered, and hence precision of the neutrino flux is limited.
One purpose of the replica target measurement is to reduce the systematic uncertainties not covered by the thin target measurements.
To reproduce the condition of neutrino oscillation experiments, proton beams with the same momentum are shot on the replica target.
For this purpose, main goal is to measure the yields of produced hadrons exiting target.

\begin{figure}[htb]
\centering
\includegraphics[height=50mm]{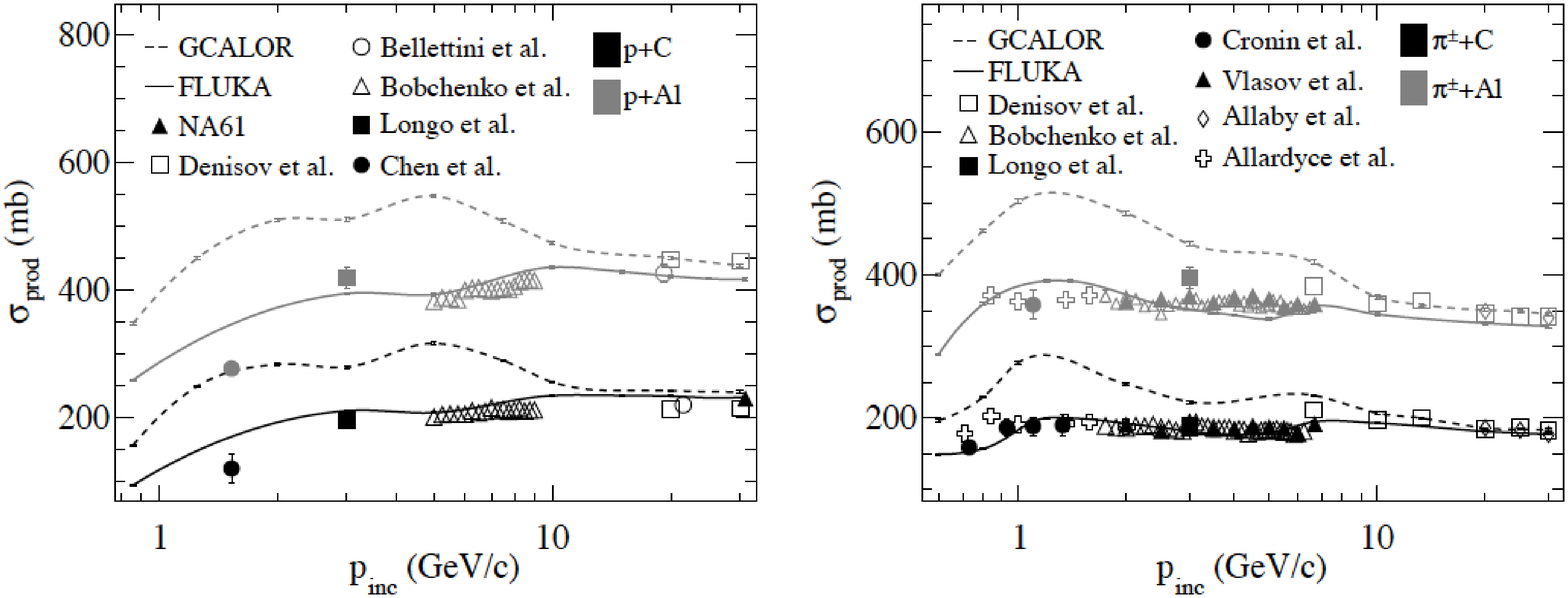}
\includegraphics[height=50mm]{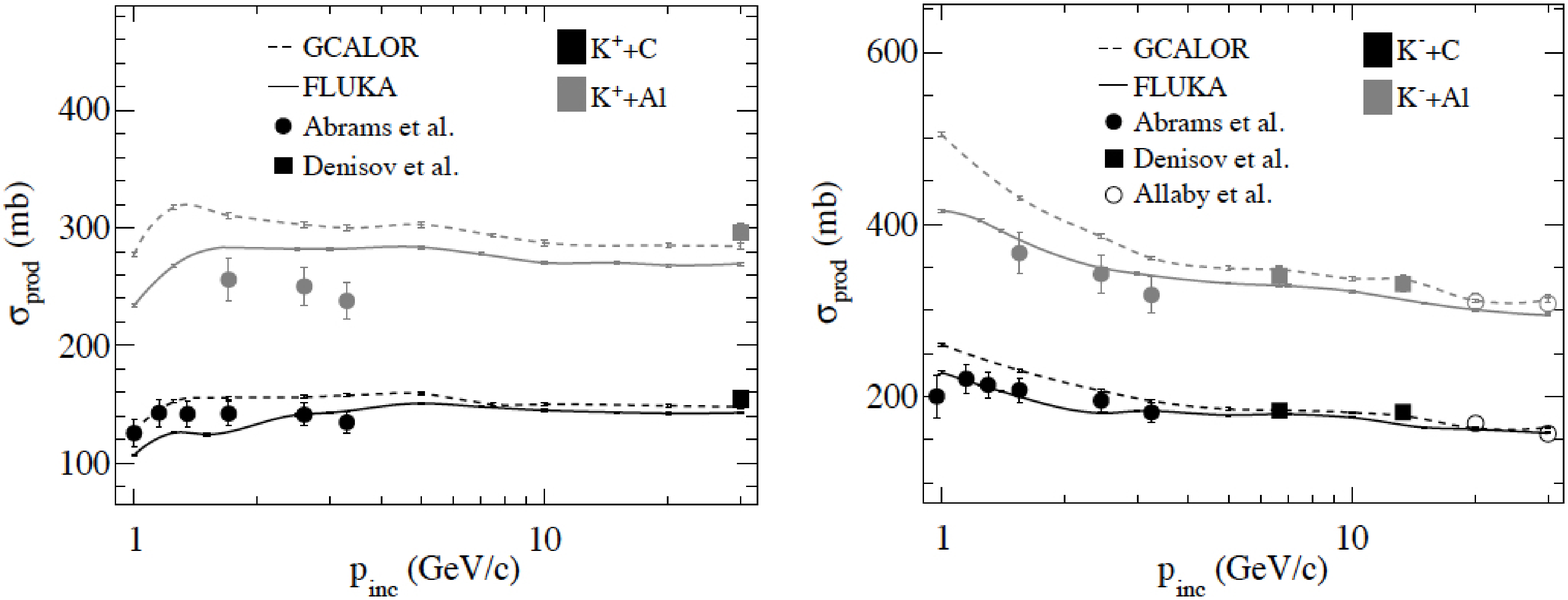}
\caption{Comparison of hadron production cross-section data with models for protons (left top), charged pions (right top), $K^+$ (bottom left), and $K^-$ (bottom right).}
\label{Fig:InelasticCrossSection}
\end{figure}

Thin and replica target measurements are complementary and have been performed by several hadron production experiments.
Figure~\ref{Fig:InelasticCrossSection} shows an example of existing hadron production cross-section measurements with thin targets~\cite{Abe:2012av}. 
In addition to these available datasets, recent experiments performed further hadron production measurements for both thin and replica targets.
The HARP experiment~\cite{Catanesi:2007ig} at the CERN PS performed hadron yield measurements using 1.5-15\, GeV/$c$ protons on various nuclear target. 
Datasets taken by HARP are mainly used for the K2K experiment and Fermilab Booster neutrino experiments.
In addition, the HARP datasets are also used for the atmospheric neutrino flux prediction.
The MIPP experiment~\cite{Raja:2005sh} at the Fermilab main injector performed hadron production measurements using 120\,GeV/$c$ primary protons and 5-90\,GeV/$c$ secondary hadrons ($\pi^{\pm}$, $K^{\pm}$, $p$, or $\bar{p}$) on various nuclear target. 
Datasets taken by the MIPP experiment are mainly used for the Fermilab NuMI beamline neutrino experiments.
The NA61/SHINE experiment at the CERN Super Proton Synchrotron (SPS) is the only running hadron production experiment at present and will be reviewed in the next section.

\section{NA61/SHINE measurements for T2K}

NA61/SHINE is a fixed-target experiment at the CERN SPS, which studies hadron production in hadron-nucleus and nucleus-nucleus collisions for various physics goals.
For neutrino physics, hadron beams (protons, pions, and kaons) are collided with a light nuclear target (carbon, aluminum, and beryllium) and spectra of outgoing hadrons are measured to improve precision of the neutrino flux prediction.
For the T2K experiment, data taking was completed by 2010 with 31\,GeV/$c$ proton beam and the analysis of the ultimate dataset is being finalized.

\begin{figure}[htb]
\centering
\includegraphics[height=66mm]{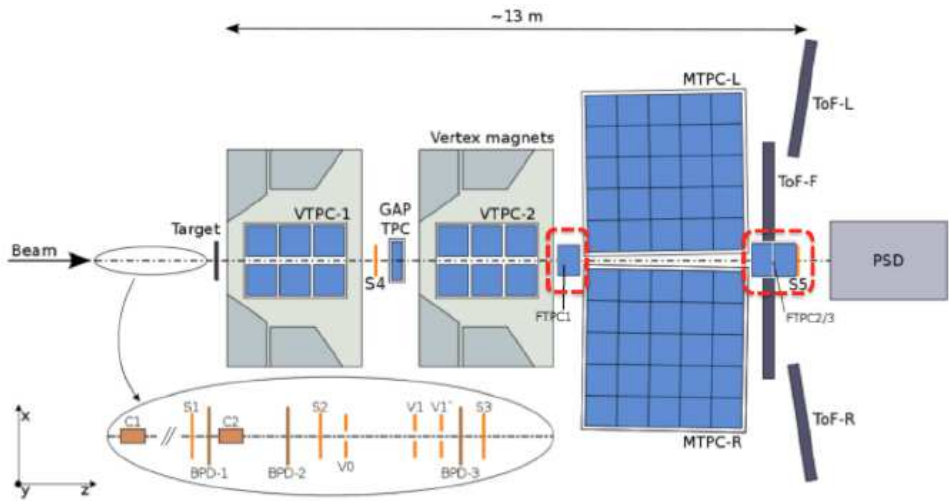}
\caption{Top view of the NA61/SHINE facility. In addition to the existing detectors, locations for the new forward TPCs are shown with dotted red line, which starts taking data from summer 2017.}
\label{Fig:NA61}
\end{figure}

\begin{figure}[htb]
\centering
\includegraphics[height=51mm]{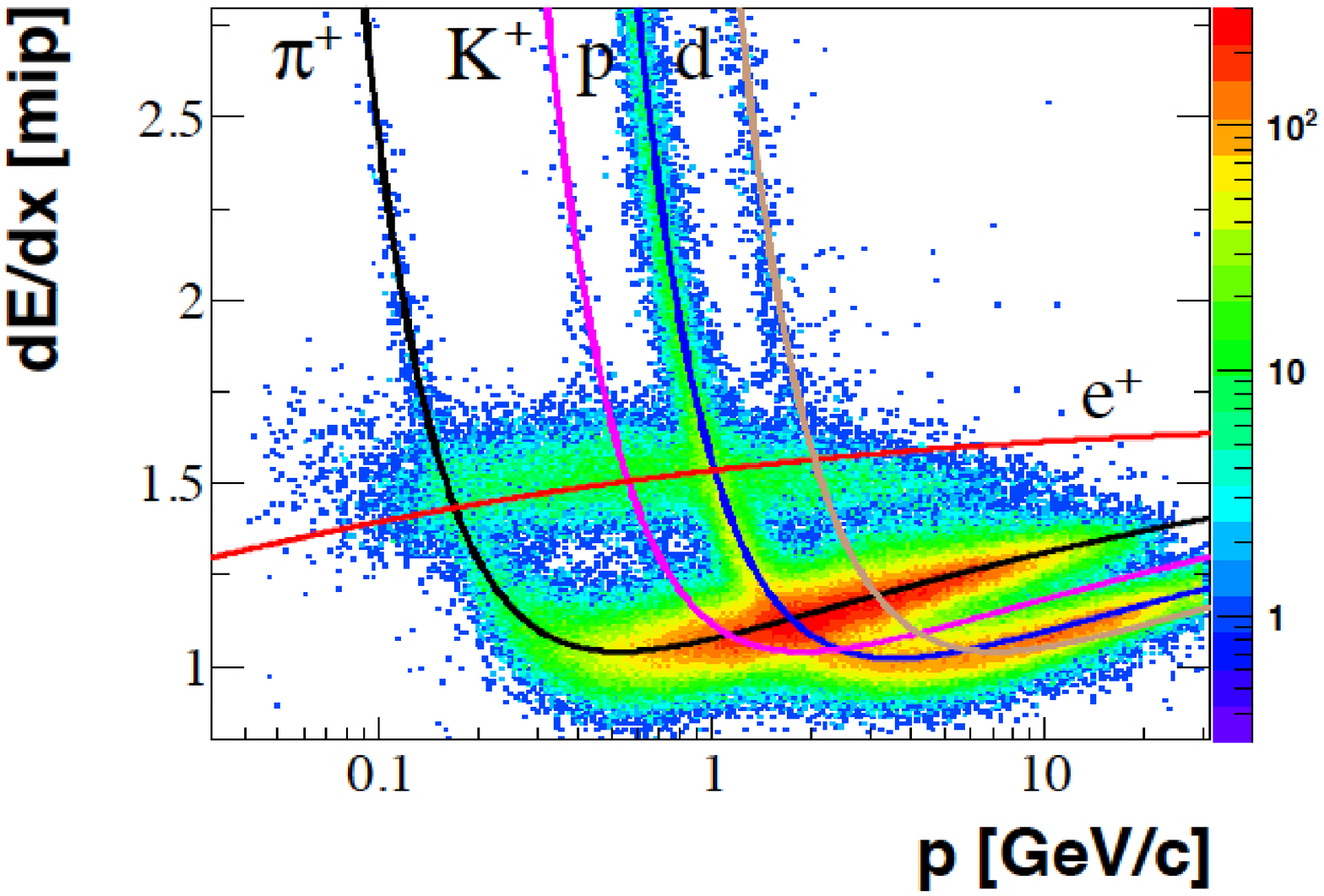}
\includegraphics[height=51mm]{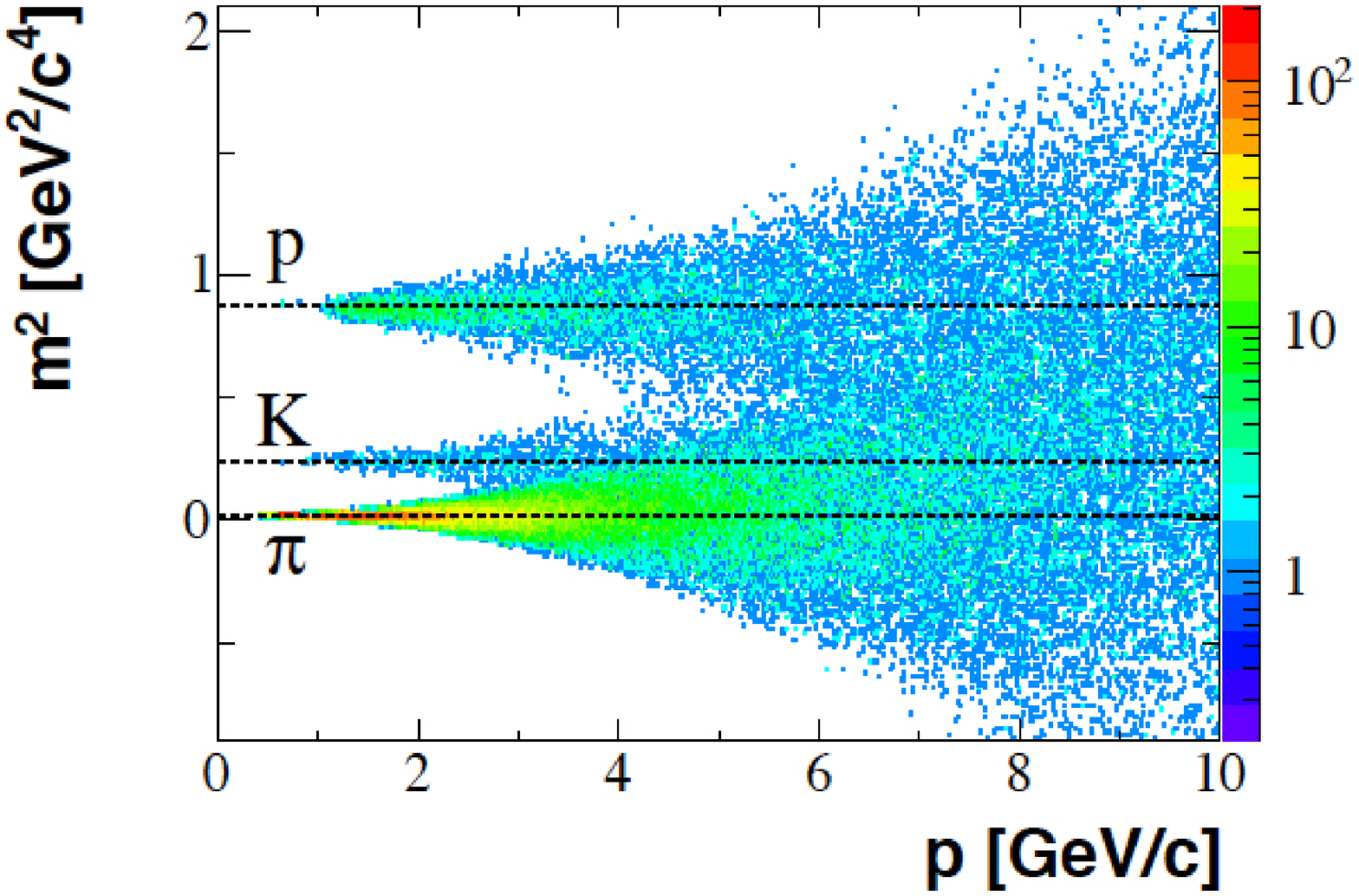}
\caption{Particle identification performance for 31\,GeV/$c$ proton beam on carbon target, as a function of particle momentum. (Left) Bethe-Bloch dE/dx curves for positively charged particles. (Right) The mass squared distribution obtained from the time of flight measurement. }
\label{Fig:NA61pid1}
\end{figure}

\begin{figure}[htb]
\centering
\includegraphics[height=58mm]{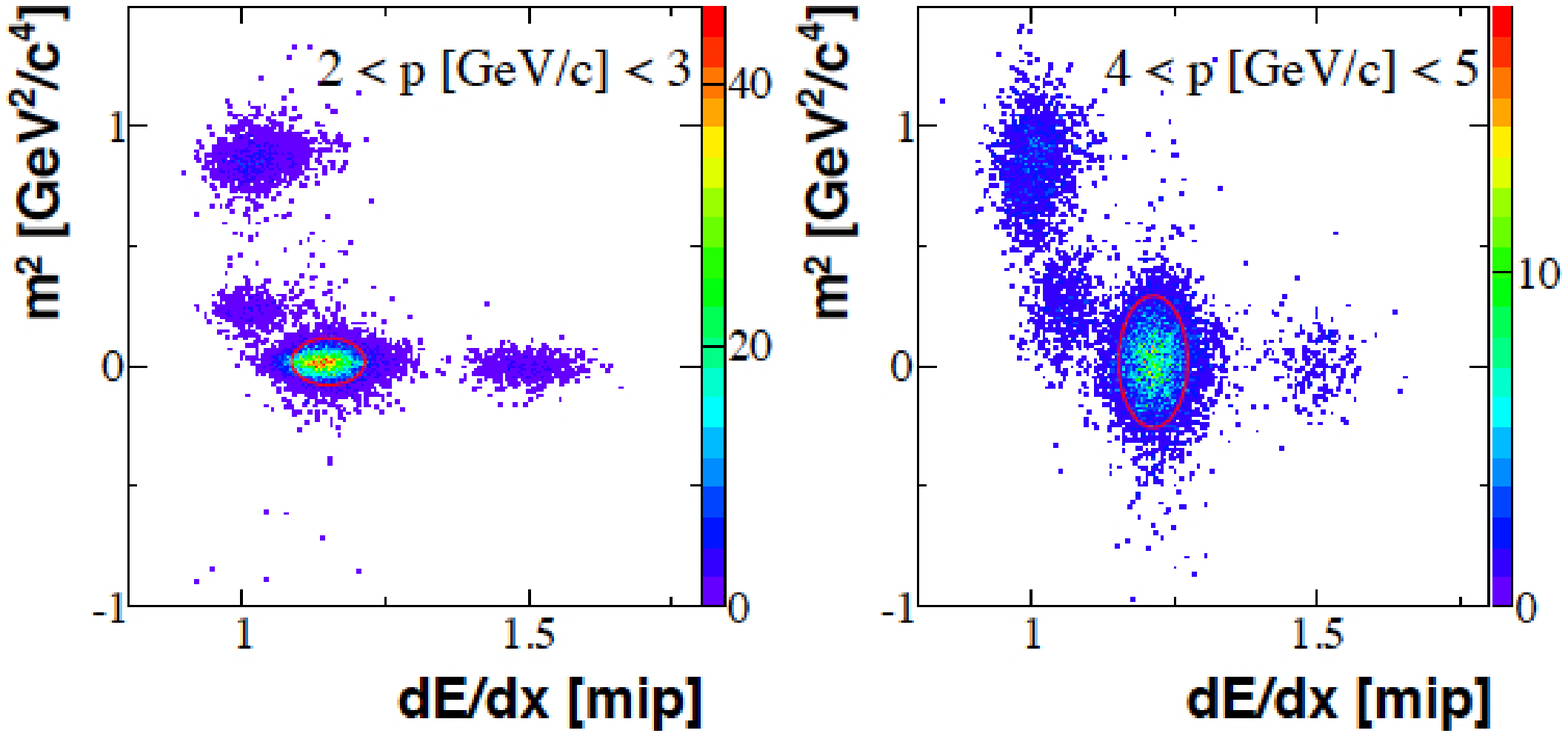}
\caption{Performance of combined dE/dx and time of flight particle identification. Charged particles in the momentum range of 2-3\,GeV/$c$ (left) and 4-5\,GeV/$c$ (right) are shown.}
\label{Fig:NA61pid2}
\end{figure}

The NA61/SHINE apparatus is a large acceptance spectrometer on the CERN SPS H2 beamline. 
Figure~\ref{Fig:NA61} shows the NA61/SHINE experimental setup. 
Main tracking detectors are four large TPCs, where
two of them sit inside the super-conducting dipole magnets 
and other two are located downstream of magnets symmetrically with respect to the beamline.
These TPCs provides good momentum reconstruction and particle identification capabilities. 
Scintillator-based time of flight detectors are located downstream of TPCs, which give complementary particle identification capability especially for the region where the Bethe-Bloch dE/dx curves overlap.
Figure~\ref{Fig:NA61pid1} shows an example of particle identification based on dE/dx and time of flight information, and its combined performance is shown in Figure~\ref{Fig:NA61pid2}~\cite{Abgrall:2011ae}.

\begin{figure}[htb]
\centering
\includegraphics[height=48mm]{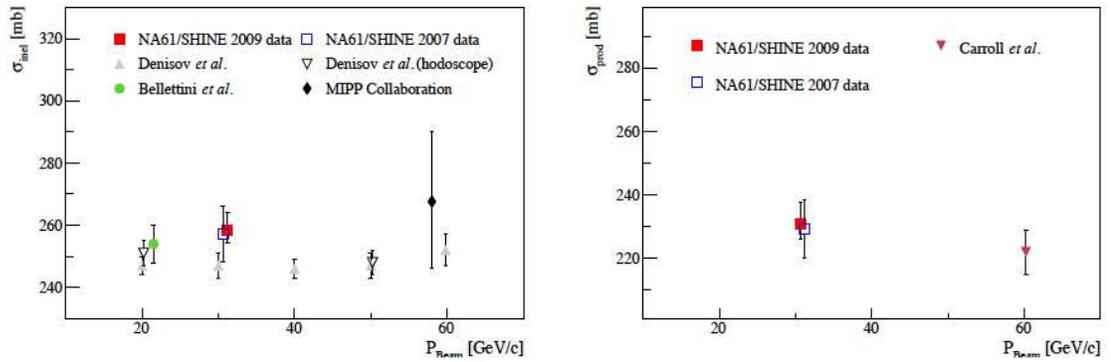}
\caption{NA61/SHINE measurement for p+C at 31\,GeV/c compared to former measurements. (Left) Inelastic cross-section. (Right) Production cross-section.}
\label{Fig:NA61inelsigma}
\end{figure}

\begin{figure}[htb]
\centering
\includegraphics[height=50mm]{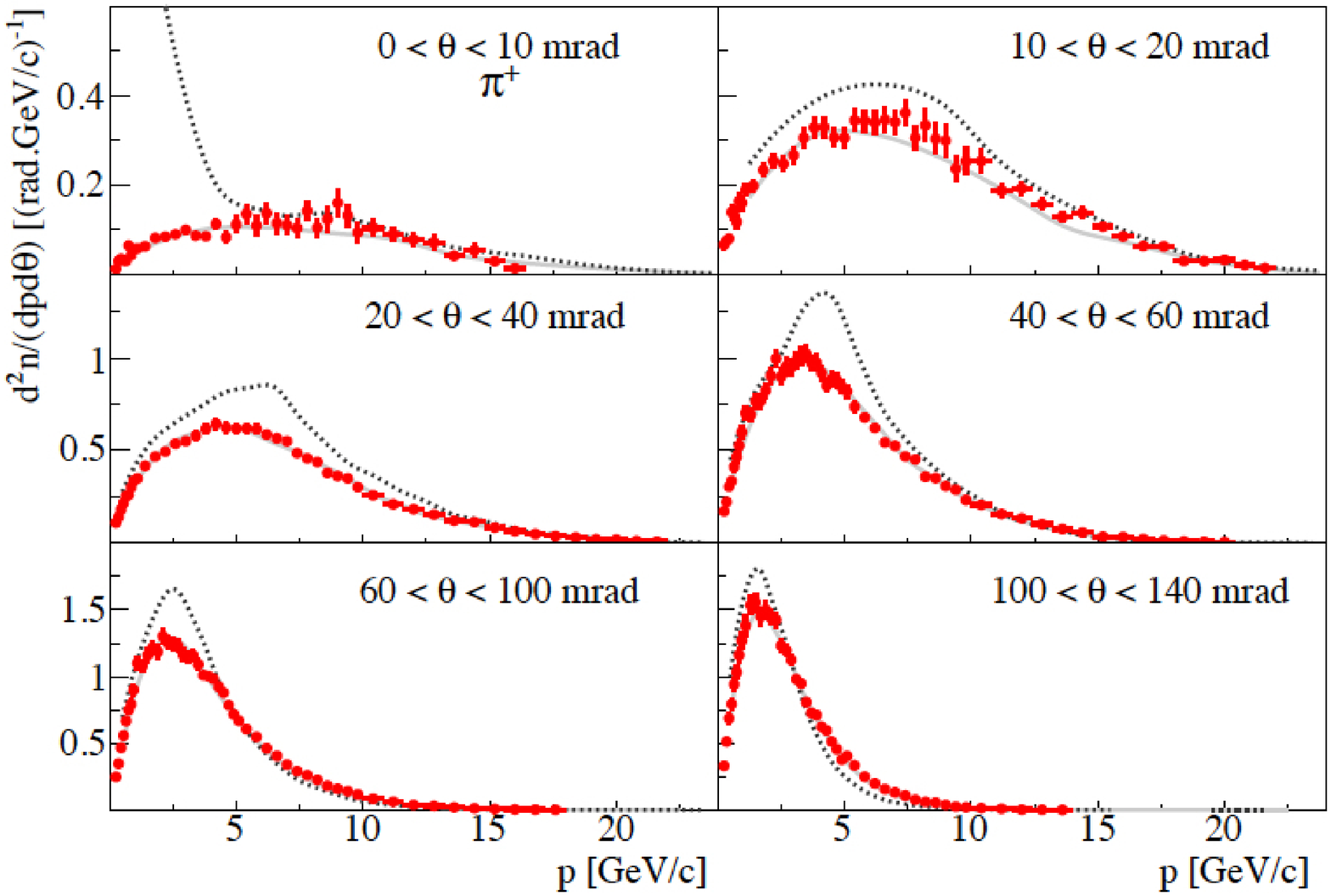}
\includegraphics[height=50mm]{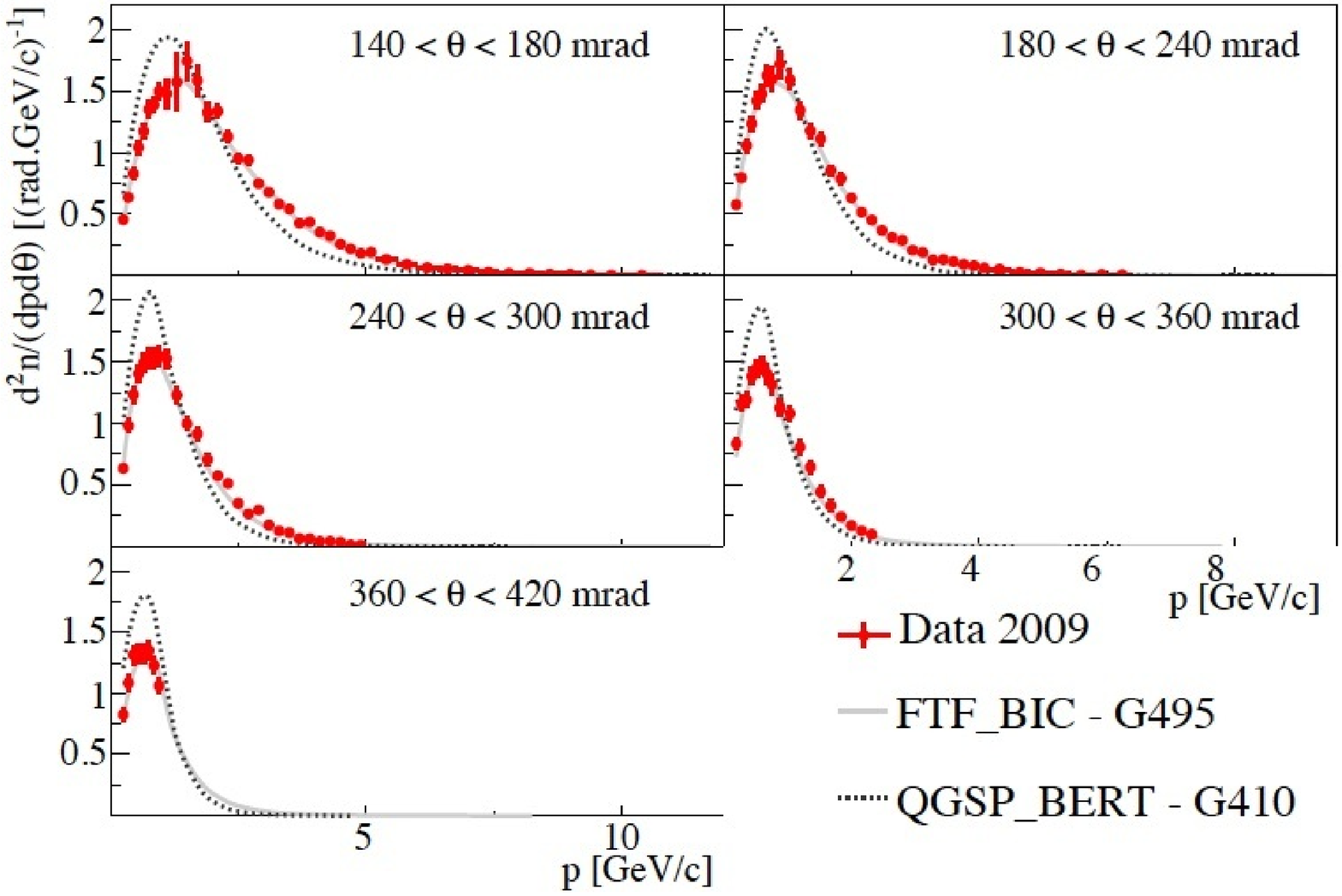}
\caption{Momentum distributions of $\pi^+$ mesons produced in p+C interactions at 31\,GeV/$c$ in different polar angle intervals. NA61/SHINE data is compared with predictions of FTF\_BIC in Geant4.9.5 and QGSP\_BERT in Geant4.10 models.}
\label{Fig:NA61piplus}
\end{figure}

NA61/SHINE published a hadron production measurement with 31\,GeV/$c$ proton beam on the thin carbon target ($\sim$0.04\,$\lambda_{\rm int}$, denote as p+C)~\cite{Abgrall:2015hmv}.
Inelastic and production cross-sections were measured with high precision and found good agreement compared to former 
measurements (Figure~\ref{Fig:NA61inelsigma}).
In addition, spectra of $\pi^\pm$, $K^\pm$, $p$, $K^0_S$, and $\Lambda$ were measured and compared with various hadron production models. 
As an example, Figure~\ref{Fig:NA61piplus} shows $\pi^+$ spectra compared with predictions by two hadron production models.

\begin{figure}[htb]
\centering
\includegraphics[height=50mm]{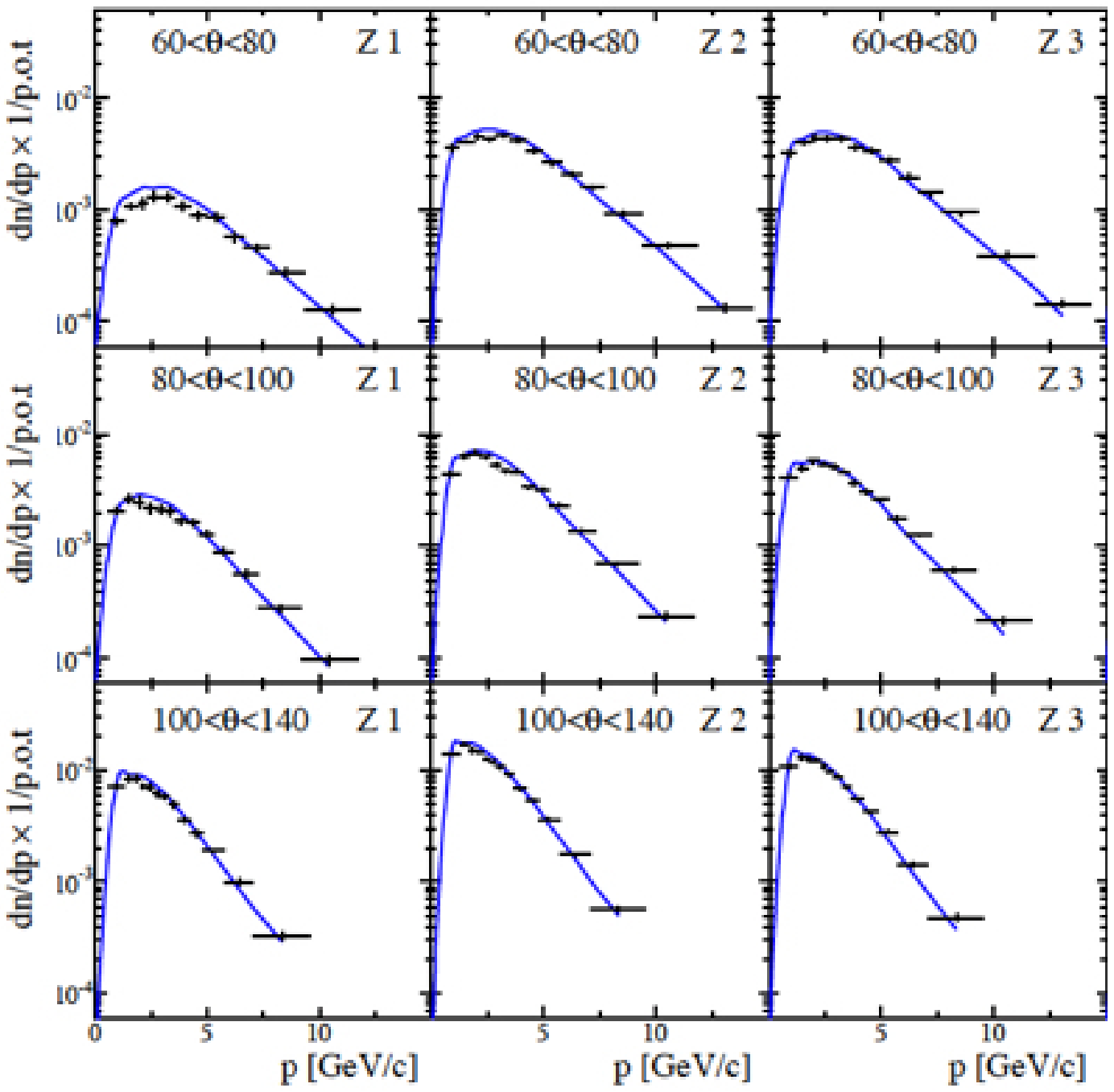} \,\,\,
\includegraphics[height=50mm]{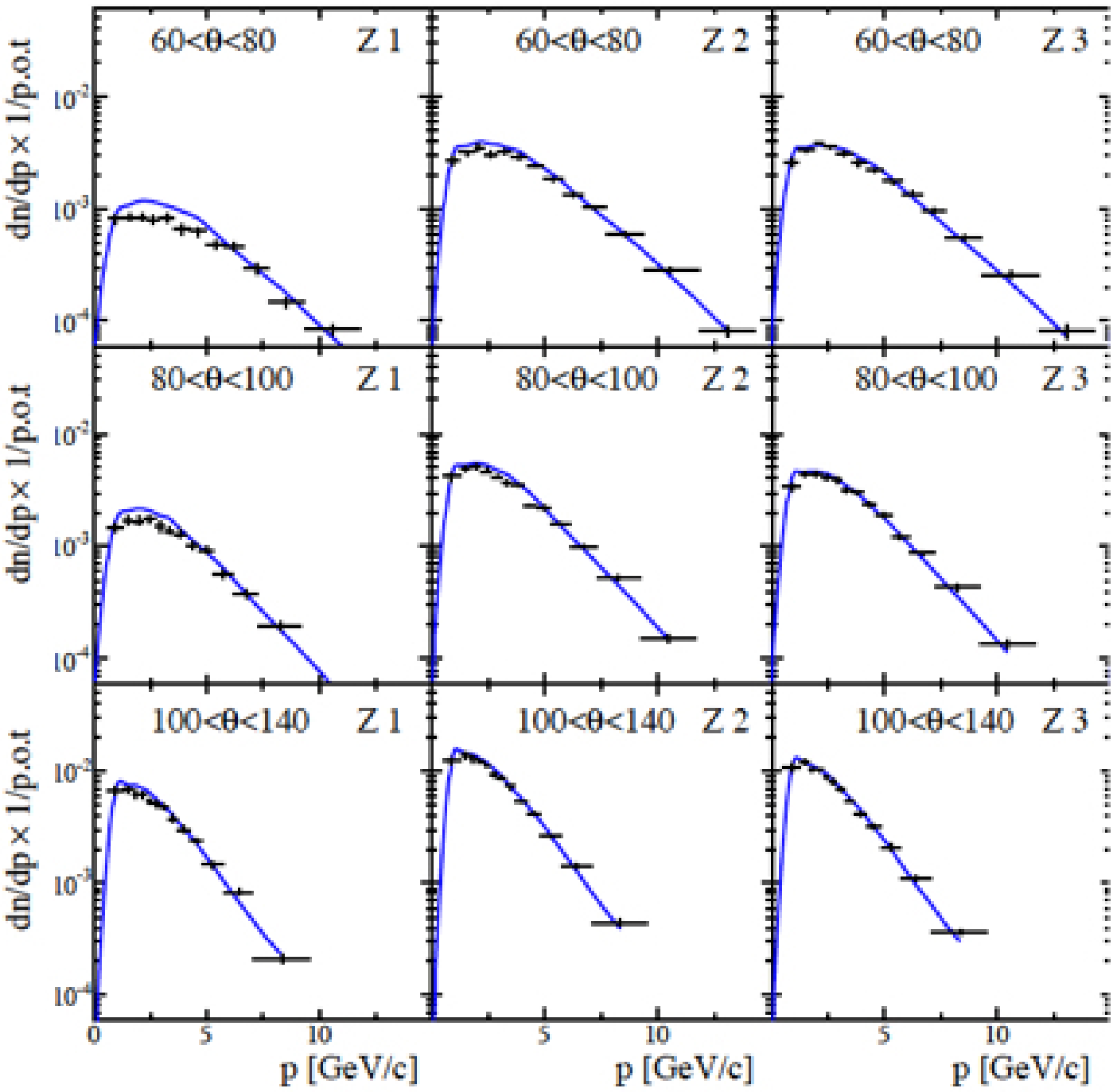}
\caption{Spectra of $\pi^+$ (left) and $\pi^-$ (right) mesons at the surface of the T2K replica target produced in p+C interactions at 31\,GeV/$c$. NA61/SHINE data (black point) is compared with the FLUKA2011 predictions (blue solid line). }
\label{Fig:NA61T2Kreplica}
\end{figure}

NA61/SHINE published their measurements with the T2K replica target more recently~\cite{Abgrall:2016jif}.
Yields of $\pi^\pm$ mesons from the surface of the T2K replica target were measured and compared with the FLUKA2011 prediction.
Figure~\ref{Fig:NA61T2Kreplica} shows an example of observed differential yields of $\pi^+$ mesons.
In addition, the neutrino flux re-weighted with this T2K replica target measurement was compared with the FLUKA2011 prediction re-weighted with the NA61/SHINE thin target cross-section measurement and found a good agreement between two measurements.

\begin{figure}[htb]
\centering
\includegraphics[height=49mm]{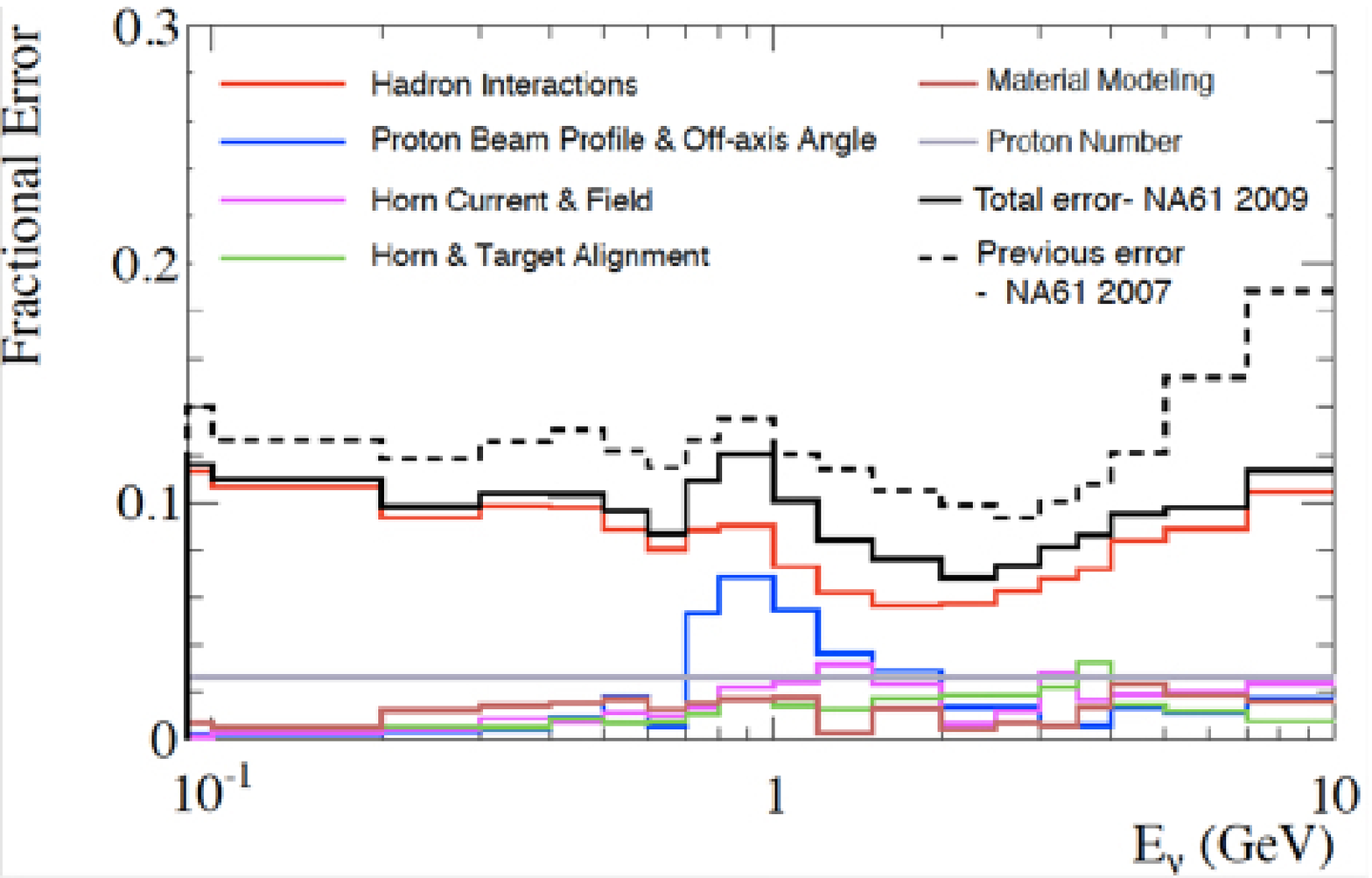}
\includegraphics[height=49mm]{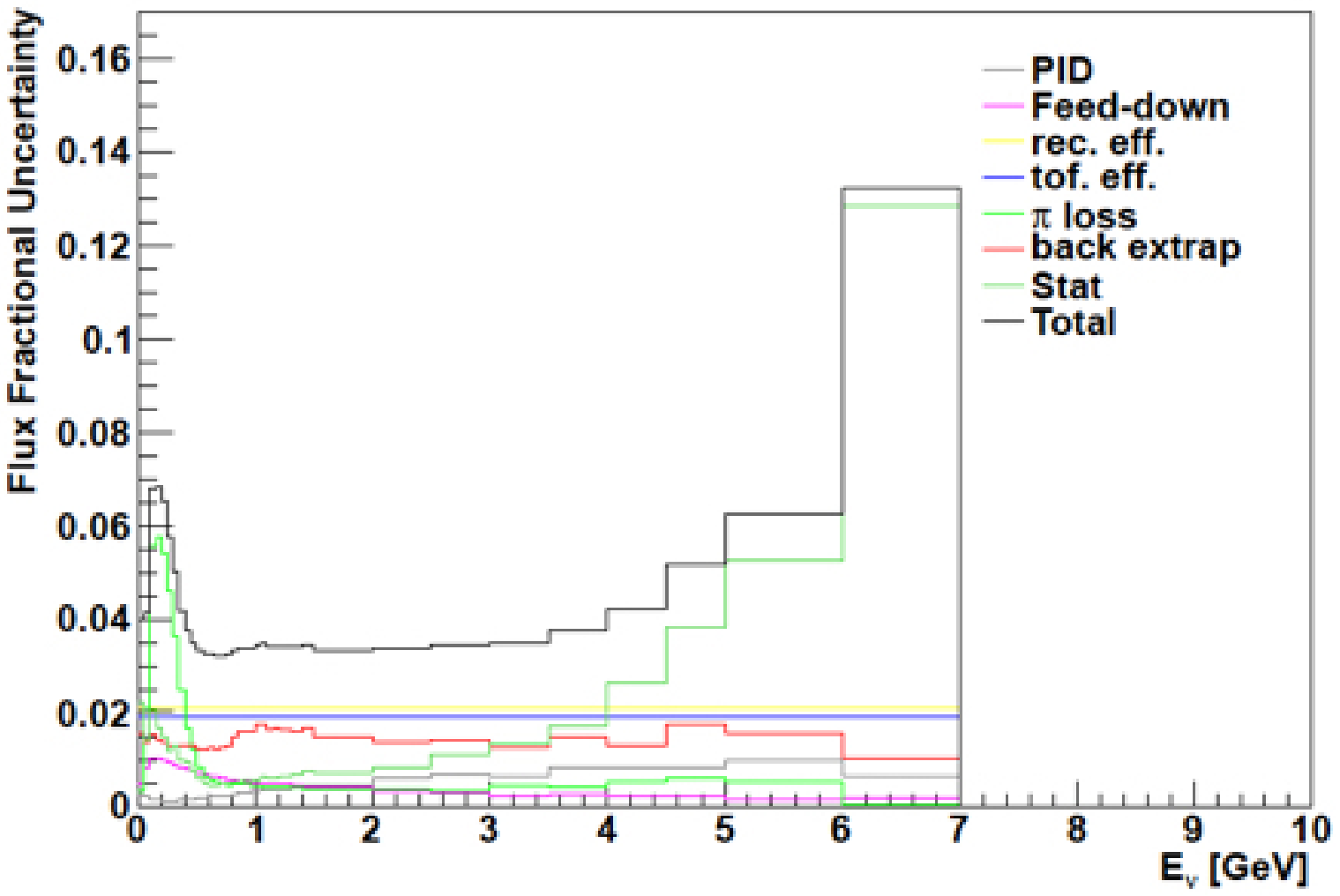}
\caption{(Left) Systematic uncertainty on the T2K neutrino flux prediction before and after applying the latest NA61/SHINE thin target measurement. (Right) Systematic uncertainty on the T2K neutrino flux prediction with the replica target measurement.}
\label{Fig:T2KFluxUnc}
\end{figure}

With the NA61/SHINE thin target measurement on p+C interactions, neutrino flux uncertainty at T2K has been successfully reduced about 25\% compared to the previous flux uncertainty (Figure~\ref{Fig:T2KFluxUnc}, left).
In addition to the thin target measurement, it was demonstrated that the neutrino flux uncertainty from the pion production contribution can be constrained up to 4\% level with the published replica target measurement (Figure~\ref{Fig:T2KFluxUnc}, right).
Improved results will be obtained from four times higher statistics dataset on the T2K replica target taken by NA61/SHINE in 2010 and further flux uncertainty reduction is expected.

\section{Future prospects}

The next generation neutrino oscillation experiments are being proposed, such as LBNF/DUNE 
and successor experiments of T2K (T2K Phase 2 and Hyper-Kamiokande experiments).
The LBNF beamline will shoot protons with momentum somewhere between 60 and 120\,GeV/$c$ on carbon or beryllium target, similarly as the NuMI beamline with 120\,GeV/$c$ protons. 
Because the primary protons have high momentum, secondary protons tend to be produced beam-forward direction and their re-interactions contribute a lot to the neutrino flux for the LBNF and NuMI beamlines. 
Therefore, it is very important to measure forward proton productions to further improve the neutrino flux prediction.
The T2K beamline will be re-used for its successor experiments with the same beam momentum and upgraded beam intensity, and there exists a possibility to re-design the target for future operations.
For all the next generation experiments, hadron production measurements with replica targets are highly desirable once their design is fixed. 

NA61/SHINE has started taking data for the LBNF and NuMI beamlines since 2015 through 2018 with higher energy beams than for T2K.
Through 2015 and 2016, various datasets have been taken for the beam simulation tunings: p+C at 31/60/120\,GeV/$c$, p+Be at 60/120\,GeV/$c$, $\pi^+$+C/Al at 31/60\,GeV/$c$, $\pi^+$+Be at 60\,GeV/$c$, and $K^+$+C/Al at 60\,GeV/$c$.
These rich datasets have been collected by NA61/SHINE and being analyzed.
In addition, forward TPCs (FTPCs) are being installed to the NA61/SHINE facility to fill the forward direction acceptance gap of the NA61/SHINE facility (as shown in Figure~\ref{Fig:NA61}).
FTPCs will join the data taking since summer 2017 through 2018 and new data will be used to provide the precise neutrino flux prediction for the LBNF and NuMI beamlines. 

The NA61/SHINE collaboration is preparing program extension proposal after CERN Long Shutdown 2 (2019-2020). 
Significant facility modifications are planned including 
TPC electronics upgrade which significantly increases readout rate up to 1000\,Hz, 
installation of modern silicon-based tracking detectors surrounding the target which improves vertex reconstruction precision drastically,
and new time of flight detectors with high time resolution ($\sigma_{\rm time} \sim$50\,ps) which increases particle identification resolution, in addition to the existing large acceptance TPCs.
Therefore, it is a great opportunity for the next generation neutrino experiments to perform precise measurements of hadron production based on their demands at the upgraded NA61/SHINE facility.






\end{document}